\documentclass[%
 aip,
 amsmath,amssymb,
 preprint,%
]{revtex4-1}
\usepackage{graphicx}
\DeclareGraphicsExtensions{.pdf,.png,.jpg,.eps}
\usepackage{xcolor}
\usepackage{caption}
\usepackage{subcaption}

\newcommand{\gv}[1]{\ensuremath{\mbox{\boldmath$ #1 $}}} 
\newcommand{\guv}[1]{\ensuremath{\mbox{\boldmath$ \hat{#1} $}}} 

\draft 

\makeatletter
\@booleanfalse\titlepage@sw
\def\frontmatter@abstract@produce{%
  \par
  \addvspace{\frontmatter@preabstractspace}%
  \begingroup
   \dimen@\baselineskip
   \setbox\z@\vtop{\unvcopy\absbox}%
   \advance\dimen@-\ht\z@\advance\dimen@-\prevdepth
   \@ifdim{\dimen@>\z@}{\vskip\dimen@}{}%
  \endgroup
  \begingroup
   \prep@absbox
   \unvbox\absbox
   \post@absbox
  \endgroup
  \@ifx{\@empty\mini@notes}{}{\mini@notes\par}%
  \addvspace\frontmatter@postabstractspace
}%
\makeatother

\begin{document}


\title{Particle pinch in global tokamak edge simulations}



\author{Ben Zhu}
\email[]{ben.zhu@columbia.edu}
\affiliation{Lawrence Livermore National Laboratory, Livermore, CA 94550}
\affiliation{Dept. of Applied Physics and Applied Mathematics, Columbia University in the City of New York, New York, NY 10027}
\affiliation{Columbia Fusion Research Center, Columbia University in the City of New York, New York, NY 10027}



\date{\today}

\begin{abstract}
The inward particle flux, or particle pinch, is routinely observed in magnetically confined fusion experiments, yet its mechanism is not fully understood. We study the particle pinch in the tokamak edge with the flux-driven global turbulence model GDB. Starting from a flat density profile fueled only near the last closed flux surface, the simulation develops a strong inward particle flux that builds a centrally peaked density profile over $O(10)$ ms.
The pinch coexists with the usual outward turbulent heat transport; two distinct mechanisms carry it. In the early stage, when density and temperature gradients oppose each other ($\eta_\alpha=L_n/L_{T_\alpha}<0$), drift-wave turbulence drives the inward flux through electron thermal diffusion. In the late stage, once the density profile has flattened, a persistent inward equilibrium $E\times B$ flux, carried by the poloidally asymmetric density and electrostatic potential, drives the central peaking. While the Pfirsch--Schl\"uter neoclassical transport sets the amplitude of the up-down asymmetric density, classical theory predicts no net radial flux from this asymmetry at leading order~\cite{hazeltine1973collision,helander2002collisional}. The observed flux flows in a separate non-ideal channel, opened by parallel resistivity, electron inertia, and electromagnetic induction in the equilibrium electron force balance. This channel shifts the equilibrium potential poloidally against the density by $\delta_s\simeq-0.08\pi$ and supplies the late-phase density build-up.
These results show that a flux-driven global edge simulation can self-consistently produce a centrally peaked density profile without ad hoc assumptions, and bear on longstanding edge questions such as density pedestal formation.

\end{abstract}

\maketitle 


\section{Introduction}
\label{sec:intro}

Particle pinch, a process in which electrons and ions are transported up (rather than down) the plasma pressure gradient, is a fundamental and yet not fully understood process in plasma physics. A peaked density profile, which implies an inward particle flux, was experimentally observed in many magnetically confined fusion devices, including linear devices~\cite{cui2015up}, magnetic dipoles~\cite{boxer2010turbulent}, tokamaks~\cite{hoang2003particle} and stellarators~\cite{stroth1999evidence}. 
In a toroidal plasma, the evolution of density must follow the conservation law,
\begin{equation}
    \frac{\partial n}{\partial t}+\nabla\cdot \gv{\Gamma}_n=S_n,
\end{equation}
where $S_n$ is the source term, i.e., ionization of neutral particles. Without pellet or neutral beam injection, neutrals from gas puffing and/or wall recycling cannot penetrate far into the core in most modern magnetic fusion experiments, i.e., $S_n\simeq 0$ somewhat inside the last closed flux surface; therefore, a radial inward particle \textit{pinch} is required to balance the diffusive flux $-D_n\nabla n$ and maintain the steady-state, peaked density profile commonly observed in experiments, i.e.,
\begin{equation}
    \Gamma_n=-D_n\nabla n-nv_\text{pinch}.
\end{equation}
Understanding this process is crucial not only for predicting density profiles from first-principle theory but also for addressing longstanding issues in tokamak physics, such as pedestal formation during the L-H transition, density limits, and the absence of a particle transport barrier in I-mode.

Over the years, many theories have been proposed to explain the inward particle transport observed experimentally. These include the neoclassical Ware pinch, driven by a finite inductive toroidal electric field \(E_\zeta\)~\cite{ware1970pinch}, the turbulent thermal diffusion process, which is proportional to \(\nabla T_e/T_e\)~\cite{coppi1978ion}, and the turbulent equipartition due to a curved magnetic field~\cite{isichenko1995invariant}. Additionally, experimental evidence indicates that density peaking is correlated with plasma collisionality in the tokamak~\cite{angioni2003density}. Fluid~\cite{garbet2003turbulent}, linear~\cite{angioni2005collisionality}, nonlinear~\cite{angioni2009particle}, and global~\cite{estrada2006density} gyrokinetic simulations of ion temperature gradient (ITG) modes and trapped electron modes (TEM) have quantified their roles in particle transport with experimentally measured plasma parameters.

However, most of the existing studies primarily focus on the core region, with less attention given to the edge. For instance, the formation of the density pedestal in the H-mode, whether due to particle transport or fueling, remains an open question~\cite{mordijck2020overview}. This complexity arises because particle dynamics in the edge region are more intricate than in the core. The edge region is not source-free; it is essential to consider particle fueling through neutral ionization, where the fueling region, or the effective neutral penetration length, depends on local plasma parameters. Moreover, in modern tokamak experiments this particle fueling is often poloidally asymmetric~\cite{rosenthal2023inference}. Additionally, the edge region often has a steeper gradient than the core, which excites instabilities with larger amplitude fluctuations, i.e., $O(1)$. Thus, investigation of particle transport in the edge region requires flux-driven global turbulence models and running simulations over transport time scales.

On the modeling side, most edge turbulence studies over the past decade have focused on the outward plasma particle and heat transport induced by turbulence. These simulations typically concentrate on plasma dynamics, neglecting neutral dynamics, and are usually performed at a short turbulence time-scale, i.e., sub-millisecond, to investigate particular edge instability or reach turbulent saturation but without significantly depleted or relaxed plasma profile~\cite{xia2013six, fan2024theoretical}. Alternatively, many studies rely on simplified assumptions, such as fixing the core density and/or fueling particles from the core side~\cite{tamain2016tokam3x, francisquez2017global, paruta2018simulation, stegmeir2019global}. Only in recent years have edge turbulence models begun incorporating neutral dynamics alongside the plasma transport models (e.g., TOKAM3X-EIRENE~\cite{fan2019effect}, GRILLIX~\cite{zholobenko2021role}, GBS~\cite{giacomin2022gbs, mancini2023self}, SOLEDGE3X~\cite{bufferand2024global}, XGC~\cite{wilkie2024reconstruction}, and Hermes-3~\cite{dudson2024hermes}). Furthermore, long transport time-scale turbulence simulations are now becoming feasible, as demonstrated by a remarkable 90 ms-long simulation performed by SOLEDGE3X~\cite{bufferand2024global}. Nevertheless, particle transport in the edge region, in particular the inward pinch, has not yet been thoroughly studied.
Similarly, for edge transport models, where turbulent effects are approximated by effective cross-field diffusion coefficients, either the inner boundary condition for plasma density is fixed, or an effective pinch term is added to the density equation with a prescribed ad hoc pinch velocity, $v_\text{pinch}$, to maintain a density profile at the transport time scale.

Therefore, the primary question this study addresses is whether the fluid-based, flux-driven global edge turbulence model can reproduce a non-trivial plasma density profile, i.e., one that is neither hollow nor flat. If yes, what is the underlying particle pinch mechanism? If not, what physics components are missing? To explore these questions, we conducted a carefully designed edge turbulence simulation lasting approximately 10 milliseconds. The simulation revealed a robust inward particle flux that produces a centrally peaked density profile; this paper untangles the underlying pinch mechanisms. Section~\ref{sec:model} describes the physics model and simulation setup, Section~\ref{sec:result} the nonlinear results, and Sections~\ref{sec:negative_eta} and~\ref{sec:positive_eta} the pinch mechanisms of the early and late stages respectively.

\section{Physics model and simulation setup}
\label{sec:model}

The numerical model used in this study is the Global Drift Ballooning (GDB) code~\cite{zhu2018gdb} - a flux-driven, global, 3D, electromagnetic drift-reduced Braginskii-based~\cite{zeiler1997nonlinear} turbulence model that can self-consistently evolve plasma as well as flow profiles throughout the entire tokamak edge region, capturing both low frequencies ($\omega\ll\omega_{ci}$), ion scale ($k\rho_i\ll1$) turbulence and plasma equilibrium profile evolution. Six independent variables, namely plasma density $n$, electron and ion temperatures $T_{e,i}$, ion parallel velocity $v_{\parallel i}$, electrostatic potential $\phi$ and the parallel component of the perturbed magnetic vector potential $\psi$ (as $\psi=-A_\parallel$), are solved in a shift-circular magnetic geometry under the large aspect ratio assumption as follows.
\begin{align}
    \frac{\partial \ln n}{\partial t}&=-\frac{c}{B}\left[\phi,\ln n\right]-v_{\parallel i}\nabla_\parallel \ln n+\frac{2c}{B}\left(C(\phi)-\frac{C(p_e)}{en}\right)+\frac{1}{en}\nabla_\parallel j_\parallel -\nabla_\parallel v_{\parallel i}+\frac{S_n}{n},\\
    \frac{\partial \ln T_e}{\partial t}&=-\frac{c}{B}\left[\phi, \ln T_e\right]-v_{\parallel e}\nabla_\parallel \ln T_e+\frac{2}{3nT_e}\nabla_\parallel \left(\kappa_\parallel^\text{e}\nabla_\parallel T_e\right)-\frac{5}{3}\frac{2c}{eB}C(T_e) \nonumber \\
    &+\frac{2}{3}\left[\frac{2c}{B}\left(C(\phi)-\frac{C(p_e)}{en}\right)+\frac{1}{en}\nabla_\parallel j_\parallel -\nabla_\parallel v_{\parallel i}-\frac{j_\parallel}{en}\nabla_\parallel \ln n\right]+\frac{S_{T_e}}{T_e},\\
    \frac{\partial \ln T_i}{\partial t}&=-\frac{c}{B}\left[\phi,\ln T_i\right]-v_{\parallel i}\nabla_\parallel \ln T_i+\frac{2}{3nT_i}\nabla_\parallel \left(\kappa_\parallel^\text{i}\nabla_\parallel T_i\right)+\frac{5}{3}\frac{2c}{eB}C(T_i) \nonumber \\
    &+\frac{2}{3}\left[\frac{2c}{B}\left(C(\phi)-\frac{C(p_e)}{en}\right)+\frac{1}{en}\nabla_\parallel j_\parallel -\nabla_\parallel v_{\parallel i}\right]+\frac{S_{T_i}}{T_i},\\
    \frac{\partial \varpi}{\partial t}&=-\frac{2c}{eB}\left(C(p_e)+C(p_i)\right)+\nabla_\parallel\left(\frac{j_\parallel}{e}\right)-\frac{c}{3eB}C(G_i) \nonumber \\
    &-\nabla_\perp\cdot\left(\frac{nc^2}{B^2\omega_{ci}}[\phi,\nabla_\perp\phi+\frac{\nabla_\perp p_i}{en}]\right),\\
    \frac{\partial v_{\parallel i}}{\partial t}&=-\frac{c}{B}\left[\phi,v_{\parallel i}\right]-v_{\parallel i}\nabla_\parallel v_{\parallel i}-\frac{1}{m_in}\nabla_\parallel \left(p_e+p_i\right)-\frac{2}{3m_in}\nabla_\parallel G_i+\frac{2c}{eB}T_iC(v_{\parallel i}),\\
    \frac{\partial \psi^*}{\partial t}&=\frac{m_ec^2}{eB}\left[\phi,\frac{j_\parallel}{en}\right]+\frac{m_ec}{e}v_{\parallel e}\nabla_\parallel\left(\frac{j_\parallel}{en}\right)+c\nabla_\parallel \phi-\frac{c}{en}\nabla_\parallel p_e+c\eta_\parallel j_\parallel,
\end{align}
where vorticity $\varpi$ and magnetic pumping term $G_i$ are defined as
\begin{align}
    \varpi&=\nabla_\perp\cdot\frac{\langle n\rangle c}{B\omega_{ci}}\left(\nabla_\perp\phi+\frac{\nabla p_i}{en}\right),\\
    G_i&=\eta_0^i\left[\frac{c}{B}\left(C(\phi)+\frac{C(p_i)}{en}\right)-2\nabla_\parallel v_{\parallel i}\right]
\end{align}
with $\langle n\rangle$ representing the flux-surface-averaged, time-dependent density profile, i.e., a relaxed version of the Boussinesq approximation.
The modified perturbed vector potential component $\psi^*=\psi-d_e^2\nabla_\perp^2\psi$ includes the electron skin depth correction, where $d_e=c/\omega_{pe}$. 
The parallel current density $j_\parallel=\nabla_\perp^2\psi/(4\pi)$ connects electron and ion parallel velocities via the relation $v_{\parallel e}=v_{\parallel i}-j_\parallel/(en)$.
The curvature operator $C(f)=-\guv{b}_0\times\gv{\kappa}\cdot\nabla f$ is defined with magnetic line curvature $\gv{\kappa}=-\guv{b}_0\times\left(\nabla\times \guv{b}_0\right)$. The Poisson bracket is $[f,g]=\guv{b}_0\cdot\left(\nabla f\times \nabla g\right)$, and the parallel gradient $\nabla_\parallel f=\guv{b}\cdot\nabla f=\left(\guv{b}_0+\tilde{\gv{b}}\right)\cdot\nabla f\simeq\guv{b}_0\cdot\nabla f+[\psi,f]/B$ consists of both the equilibrium and perturbed magnetic field contributions.
Classical Braginskii~\cite{braginskii1965transport} transport coefficients (e.g., $\eta_\parallel$, $\kappa_\parallel^{e,i}$, $\eta_0^i$) are used in this study.  The use of collisional closure is marginally justified for the high-density, low-temperature edge plasma presented here, i.e., $n \sim 10^{20}\mathrm{m}^{-3}, T < 180\mathrm{eV}$. However, a more appropriate closure, such as the transcollisional correction~\cite{zholobenko2024tokamak}, should be implemented for weakly collisional or near-collisionless plasmas in lower-density and higher-temperature edge regions.
In addition, volumetric source terms $S_n, S_{T_{e,i}}$ are added to the density and temperature equations.
The above equations are normalized (see Appendix~\ref{app:normalization}) and solved in GDB. Similar sets of drift-reduced Braginskii transport equations are employed in other fluid-based, flux-driven tokamak edge turbulence codes, such as BOUT++~\cite{zhu2021drift}, GBS~\cite{halpern2016gbs}, GRILLIX~\cite{stegmeir2018grillix}, and Hermes-3~\cite{dudson2024hermes}.

\begin{figure}[h!]
\centering
\includegraphics[width=0.9\textwidth]{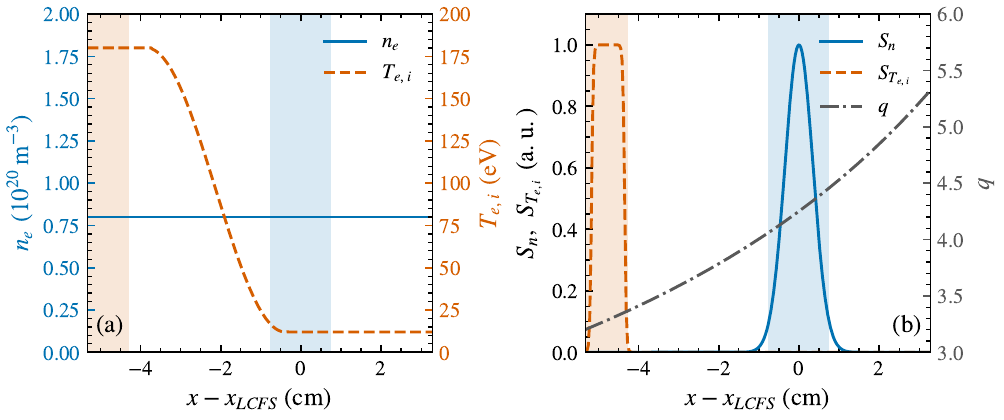}
\caption{Simulation setup: (a) initial density $n$ and temperatures $T_{e,i}$; (b) sources $S_n$, $S_{T_{e,i}}$ and safety factor $q$.  Shaded bands mark the energy (left) and particle (center) source zones here and in Figures~\ref{fig:profiles_t2_sat} and~\ref{fig:delta_s}.}
\label{fig:setup}
\end{figure}

For this study, a simulation is set up for a singly-charged deuterium plasma with Alcator C-Mod-like dimensions ($R=68.5$ cm, $a=22$ cm) and discharge parameters (e.g., $B_0=5.3$ T) in an inner-wall limited configuration. 
The simulation domain encompasses the entire toroidal annulus, radially spanning across the last closed flux surface (LCFS). The radial domain extends from about 5.3 cm inside the LCFS to about 3.3 cm in the scrape-off-layer (SOL), i.e., the normalized radial domain is $r/a\in(0.76, 1.15)$. The mesh resolution is $(n_x,n_y,n_z)=(256,1024,32)$ where $x,y,z$ represent the radial, poloidal, and toroidal directions, respectively.

To highlight the particle pinch phenomenon, a flat density profile is initialized with $n_0=0.8\times 10^{20}m^{-3}$. Particle and heat sources $S_n,S_{T_{e,i}}$ are radially localized as depicted in Figure~\ref{fig:setup}. The top-hat-shaped heat sources $S_{T_{e,i}}$ are positioned on the core side, where $x-x_\text{LCFS}<4.4$ cm or $r/a<0.8$. Their amplitude is self-regulated to maintain core side electron and ion temperatures at approximately $180$ eV. Conversely, the particle source $S_n$ is Gaussian-shaped and located at the LCFS. It is defined as $S_n=A_n\exp\left[-(x-x_\text{LCFS})^2/(2\sigma^2)\right]$ with $\sigma=0.35$ cm, so the effective source region (above $5\%$ of peak) is $|x-x_\text{LCFS}|\leq 0.87$ cm. The particle source rate, $\int dV S_n$, is set to $5.2\times10^{22}s^{-1}$ in our simulation to align with the measured ionization flux in Alcator C-Mod main chamber~\cite{labombard2001particle}, which typically ranges from 2 to $10\times 10^{22}s^{-1}$. We note that the poloidally symmetric source used here is an idealization: experimentally inferred neutral-ionization sources are themselves strongly poloidally asymmetric~\cite{rosenthal2023inference}. A poloidally localized source would modify the plasma structure within the fueling zone, but not the closed-flux-region transport mechanisms analyzed below, which operate several centimeters inside the source region.


In this shift-circular, inner-wall limited configuration, the GDB simulation domain is naturally periodic in poloidal (inside the LCFS) and toroidal directions. For the radial direction, a homogeneous Neumann boundary condition, $\partial f/\partial x=0$, is applied to density $n$ and temperatures $T_{e,i}$ at both inner and outer boundaries, allowing the plasma profile to evolve freely.   
In addition, the condition $\partial \phi_0/\partial x=0$ is applied at the inner boundary, where $\phi_0$ is the zonal component of the electrostatic potential $\phi$, to eliminate external particle, angular momentum, and heat fluxes from the core side.
In the SOL, Bohm sheath boundary conditions are applied at the limiter surfaces to ensure proper plasma exhaust in this flux-driven system.

\section{Nonlinear simulation results}
\label{sec:result}

We ran the simulation with the setup of the previous section, turning on the particle source $S_n$ after turbulence onset at $t_1=0.13$ ms and continuing for $8.40$ ms until the density profile reached a quasi-steady state. This transport-timescale simulation reproduces several familiar edge plasma features: the in-out asymmetric turbulence due to ballooning mode~\cite{zhu2017global}, the up-down asymmetric pattern due to Pfirsch-Schl\"uter transport~\cite{zhu2018up}, spontaneous formation of poloidal $E\times B$ flow in the electron diamagnetic direction in the closed-flux region and with a reversed flow direction in the SOL due to sheath boundary~\cite{zhu2017global,zholobenko2021electric}. 

\begin{figure}[h!]
\centering
\includegraphics[width=\textwidth]{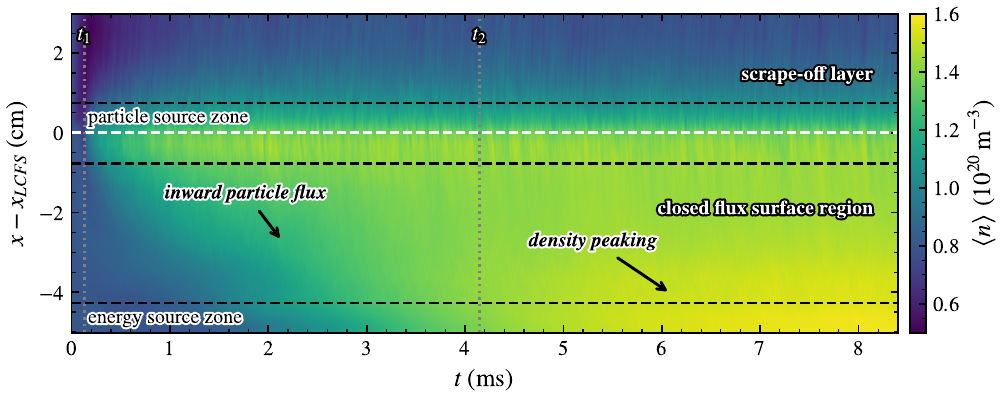}
\caption{Evolution of the flux-surface-averaged density $\langle n\rangle$.  $S_n$ turns on at $t_1=0.13$~ms; $\langle n\rangle$ inside the LCFS is roughly flat at $t_2=4.15$~ms.}
\label{fig:dvt}
\end{figure}

\begin{figure}[h!]
\centering
\includegraphics[width=0.8\textwidth]{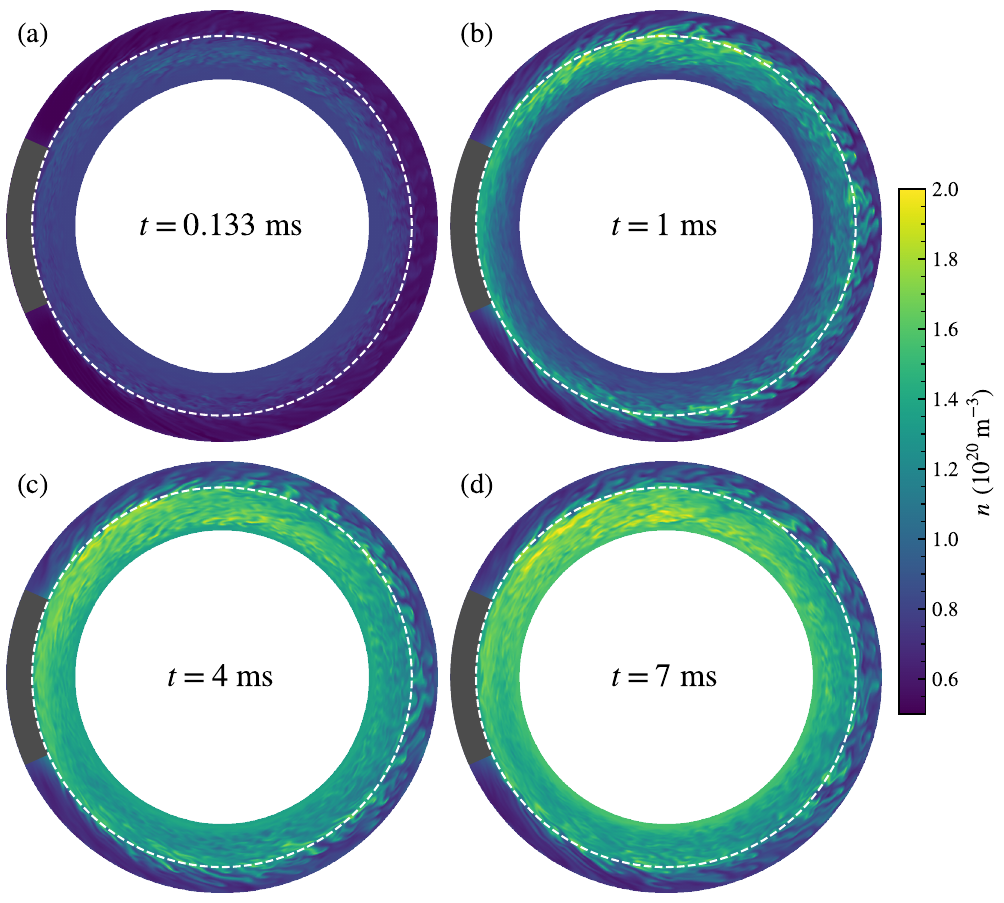}
\caption{Poloidal density cross-sections at (a) $t=0.133$~ms (before fueling), (b) $1$~ms, (c) $4$~ms ($\langle n\rangle$ nearly flat), and (d) $7$~ms (centrally peaked).}
\label{fig:den_snapshots_peaking}
\end{figure}

\begin{figure}[h!]
\centering
\includegraphics[width=0.9\textwidth]{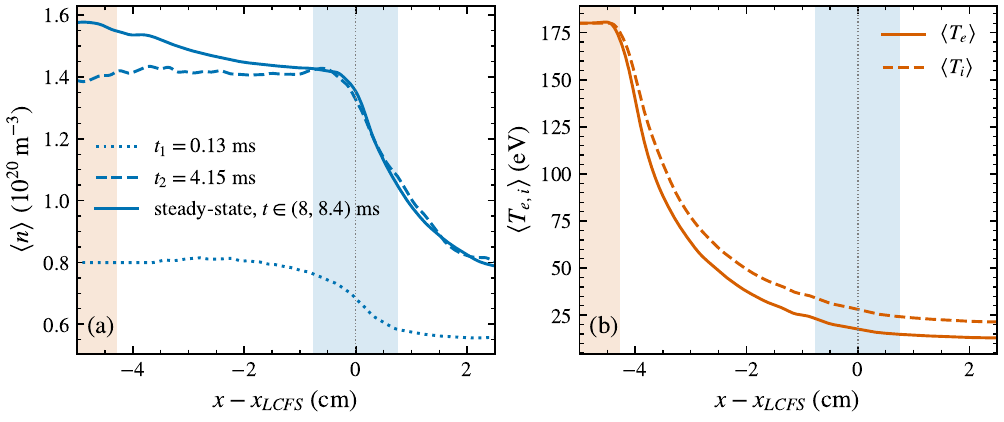}
\caption{Flux-surface-averaged profiles: (a) $\langle n\rangle$ at $t_1$ (dotted), $t_2$ (dashed), and over the steady-state window $t\in(8,8.4)$~ms (solid); (b) steady-state $\langle T_e\rangle$ (solid) and $\langle T_i\rangle$ (dashed).}
\label{fig:profiles_t2_sat}
\end{figure}

Figure~\ref{fig:dvt} shows the evolution of the radial density profile $\langle n\rangle$. Here, $\langle ...\rangle$ represents the flux-surface average (or roughly poloidal and toroidal average in shift-circular configuration) throughout the paper. Before the particle source $S_n$ was turned on, the plasma density in the SOL and near the LCFS dropped due to plasma exhaust at the limiter surfaces. Once the particle source is on, an inward particle flux emerges inside the LCFS, slowly (compared with the turbulence time-scale $O(10^{-6})$ s) ramping up the plasma density outside the particle source zone. After $4$ ms, the radial density profile is roughly flat (at $t_2=4.15$ ms). Particles continue to penetrate inwards and eventually reach a centrally peaked quasi-steady-state density profile after $8$ ms.
Throughout this evolution, the edge plasma is strongly turbulent and exhibits strong poloidal asymmetry, as indicated by selected poloidal cross-section snapshots in Figure~\ref{fig:den_snapshots_peaking}. The corresponding flux-surface-averaged profiles are collected in Figure~\ref{fig:profiles_t2_sat}: the density profile at the flat-profile time $t_2$ and in the steady-state window, and the steady-state electron and ion temperature profiles, which decrease monotonically from the heated core side ($T_{e,i}\simeq 180$~eV) to the LCFS ($T_e\simeq 18$~eV, $T_i\simeq 28$~eV). The temperature gradients thus remain negative throughout the closed-flux region during the entire evolution, so the sign changes of $\eta_\alpha$ discussed below are carried entirely by the density gradient.

The inward particle flux and density peaking confirm that a global edge turbulence simulation can produce realistic edge density profiles without ad hoc assumptions (e.g., fixed core density, core-side fueling, or a prescribed ``effective'' pinch velocity). Although the drift-reduced Braginskii description is less comprehensive than gyrokinetic and kinetic descriptions, in this collisional regime it contains the essential particle-pinch physics and therefore offers a natural starting point for studying inward particle transport in the edge.

The temporal evolution of the density profile can be divided into two stages depending on the directions of the density and temperature gradients in the closed flux surface region. Based on the simulation setup, electron and ion temperatures are always peaked on the core side, i.e., their gradients are always negative. However, the combination of initial constant density and localized particle fueling results in a positive density gradient during the early stage of the simulation, i.e., $t_1<t<t_2$. Thus, $\eta_\alpha=L_n/L_{T_\alpha}=\partial_x (\ln T_\alpha) /\partial_x (\ln n)$ is negative in this stage. Once the density profile inside the LCFS reaches a flat profile at $t_2$ and starts to peak on the core side, the density gradient becomes negative, and $\eta_\alpha$ becomes positive. 

The next two sections identify a distinct pinch mechanism in each stage.

\section{Inward particle flux at the early phase}
\label{sec:negative_eta}

This section focuses on the negative $\eta_\alpha$ phase in the simulation, i.e., $t\in(0.13, 4.15)$ ms. 
While most tokamak experiments typically exhibit a positive $\eta_\alpha$ in the edge plasma, a negative $\eta_\alpha$ is still possible. For example, a density hump can arise when the plasma is excessively fueled in the edge region, such as through massive gas puffing or pellet injection. In this simulation, we find that when $\eta_\alpha$ is negative, drift-wave instabilities coexist with resistive ballooning modes, resulting in an inward particle flux driven by the electron thermal-diffusion effect.


\subsection{Linear analysis of a simplified slab model}
To identify the instabilities and their roles in particle transport, we derived a simplified electrostatic model for a curved slab from the full set of drift-reduced Braginskii equations.
\begin{align}
    \frac{\partial}{\partial t}n&=-[\phi,n]+\epsilon_j\nabla_\parallel j_\parallel,\\
    \frac{\partial}{\partial t}T_e&=-[\phi,T_e]+\kappa_\parallel^e\nabla_\parallel^2T_e,\\
    \frac{\partial}{\partial t}T_i&=-[\phi,T_i],\\
    \frac{\partial}{\partial t}\varpi&=-C(p_e+p_i)+\nabla_\parallel j_\parallel,\\
    0&=\frac{1}{\alpha_m}\left(\nabla_\parallel\phi-\alpha_{de}\frac{\nabla_\parallel p_e}{n_0}\right)+\eta_\parallel j_\parallel
\end{align}
with
\begin{equation}
    \varpi=n_0\nabla_\perp^2\phi+\alpha_{di}\nabla_\perp^2p_i
\end{equation}

This minimal model retains four independent variables and captures the most dominant physical processes on the right-hand side of the equations. It incorporates only resistive ballooning modes (RBMs) and drift-wave instabilities (DWs), and is applicable to the source-free closed flux region. Additionally, the linearized equations are as follows:
\begin{align}
    -i\omega\tilde{n}&=ik_yn_0'\tilde{\phi}+ik_\parallel\epsilon_j j_\parallel,\\
    -i\omega\tilde{T}_e&=ik_yT_{e0}'\tilde{\phi}-k_\parallel^2\kappa_\parallel^e\tilde{T}_e,\\
    -i\omega\tilde{T}_i&=ik_yT_{i0}'\tilde{\phi},\\
    -i\omega\tilde{\varpi}&=-ik_y\epsilon_p\left[\tilde{n}\left(T_{e0}+T_{i0}\right)+n_0\left(\tilde{T}_e+\tilde{T}_i\right)\right]+ik_\parallel j_\parallel,\\
    \alpha_m\eta_\parallel j_\parallel&=ik_\parallel\left(\alpha_{de}\frac{T_{e0}}{n_0}\tilde{n}+\alpha_{de}\tilde{T}_e-\tilde{\phi}\right)\label{eq:linear_ohm}
\end{align}
with
\begin{equation}
    \tilde{\varpi}=-k_y^2\left(n_0\tilde{\phi}+\alpha_{di}\tilde{n}T_{i0}+\alpha_{di}n_0\tilde{T}_i\right).
\end{equation}
Here $n'=\partial n/\partial x$ is the drift-wave drive and the coefficient $\epsilon_p$ is used to easily turn on ($\epsilon_p=1$) and off ($\epsilon_p=0$) the curvature term.

With these linearized equations, a quadratic dispersion relation can be derived either analytically or solved numerically and the $E\times B$ particle flux in Fourier representation can be estimated by the following formula:
\begin{equation}
    \Gamma_n=\langle\tilde{n}\tilde{v}_{E,r}\rangle =-\frac{2c}{B}\mathbf{Im}\left(\Sigma_{k}k_y\tilde{n}\tilde{\phi}^*\right).
\end{equation}
Note that Linear theory determines the growth rate and the cross-phase $\delta_t$ between $\tilde{n}$ and $\tilde{\phi}$ for each mode $(k_y,k_\parallel)$, whereas the saturated amplitude estimate requires a separate quasilinear saturation rule~\cite{staebler2024quasilinear}.  The flux spectra in this section are therefore evaluated at unit mode amplitude, $\Gamma_k=-2k_y\rho_s\,\mathrm{Im}[\hat n\hat\phi^*]$ with $|\hat\phi|=1$, and carry phase and spectral-shape information only.
The range of interest is $k_y\rho_s\in(0.1,1)$.  The upper bound is set by the drift ordering which assumes $k_\perp\rho_i\ll1$, so the model is not quantitatively reliable above it; while the lower bound is set instead by the local approximation below with this approximation is no longer valid.  

\begin{figure}[h!]
\centering
\includegraphics[width=\textwidth]{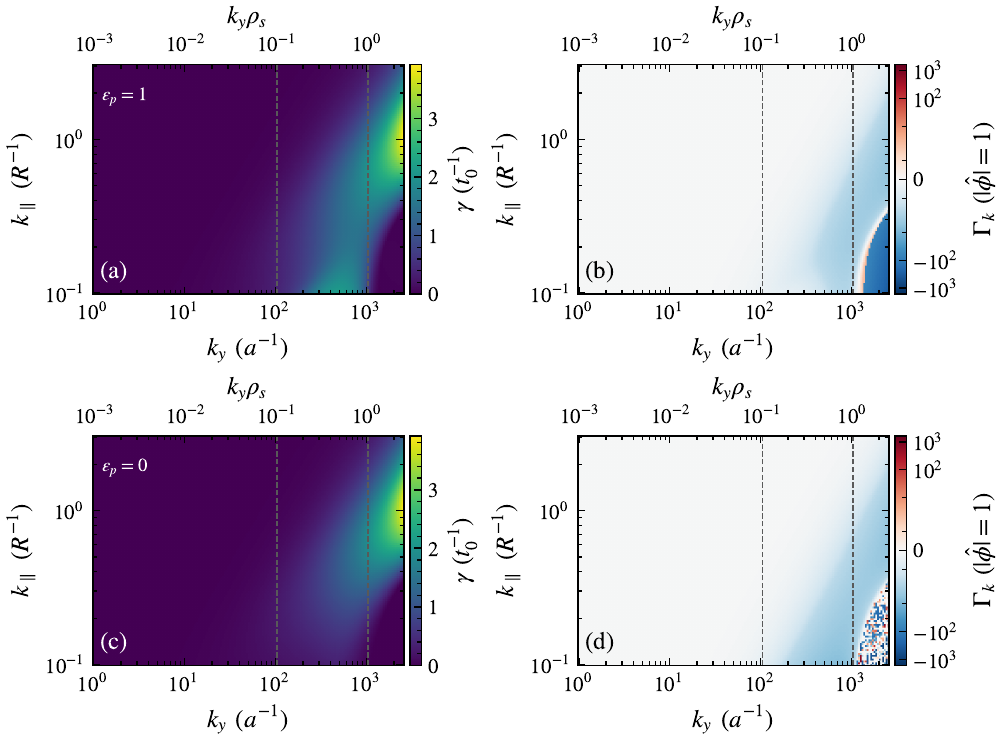}
\caption{Linear growth rate (a, c) and $E\times B$ particle flux (b, d) spectra, with (top) and without (bottom) the curvature drive, for $\eta_\alpha=-2$ ($L_n=4$ cm, $L_{T_e}=L_{T_i}=-2$ cm).  Both rows share one color scale; the shaded band marks $k_y\rho_s\in(0.1,1)$.  Same conventions in Figures~\ref{fig:lgr_eta2}, \ref{fig:lgr_woTe} and~\ref{fig:lgr_eta_inf}.}
\label{fig:lgr_eta-2}
\end{figure}
\begin{figure}[h!]
\centering
\includegraphics[width=\textwidth]{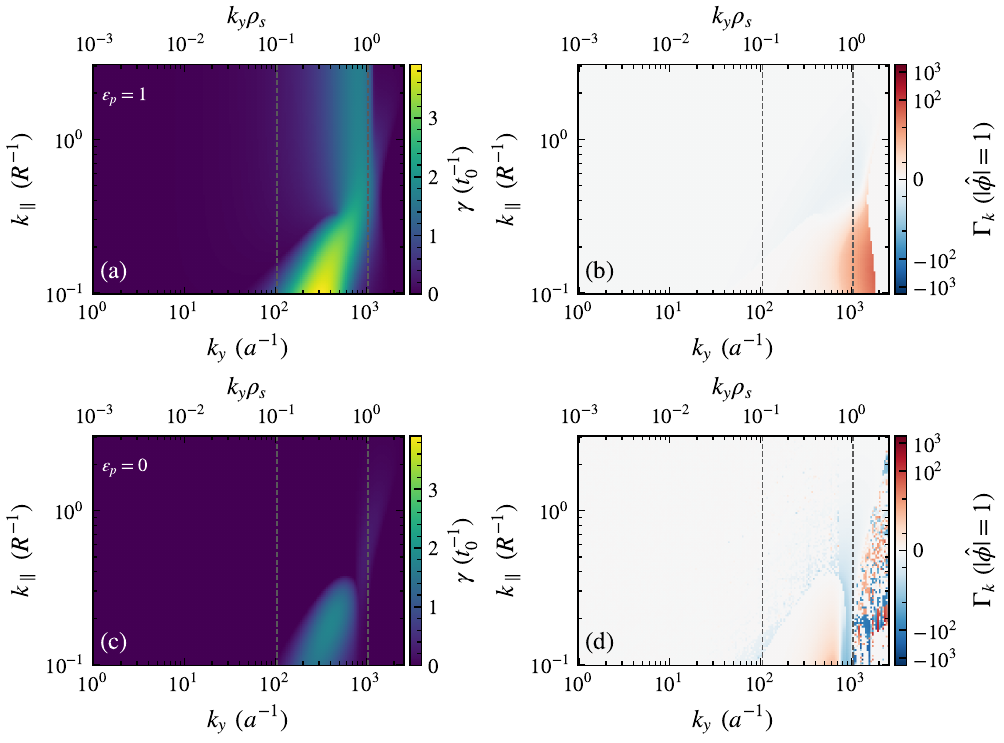}
\caption{Same setup as Figure~\ref{fig:lgr_eta-2}, but for $\eta_\alpha=2$ ($L_n=-4$ cm).}
\label{fig:lgr_eta2}
\end{figure}

Assuming the background plasma has a density $n_0=0.8\times10^{20}m^{-3}$ and temperatures $T_{e0}=T_{i0}=60eV$, the radial characteristic lengths are $L_{T_e}=L_{T_i}=-2cm$ and $L_n=4cm$. Therefore, $\eta_e=\eta_i=-2$. Solving the linearized system with these values yields the linear growth rate and the $E\times B$ particle flux contribution of each mode $(k_y,k_\parallel)$, as illustrated in Figure~\ref{fig:lgr_eta-2}. To separate drift waves from ballooning (interchange-driven) modes, we performed the calculations both with ($\epsilon_p=1$) and without ($\epsilon_p=0$) the curvature term; and to highlight the effect of negative $\eta_\alpha$, Figure~\ref{fig:lgr_eta2} shows the corresponding positive-$\eta_\alpha$ results ($L_n=-4$~cm, $\eta_e=\eta_i=2$).

Comparing Figures~\ref{fig:lgr_eta-2} and~\ref{fig:lgr_eta2}, it is evident that resistive ballooning modes (RBMs) with $k_\parallel\simeq 0$ and drift-wave instabilities (DWs) with finite $k_\parallel$ can coexist in both positive and negative $\eta_\alpha$ scenarios within this simplified linear stability analysis. However, unstable DWs exhibit a finer mode structure (higher $k_\parallel$ and $k_y$) when $\eta_\alpha<0$ compared to when $\eta_\alpha>0$, and they also have larger linear growth rates than RBMs.
More interestingly, both RBMs and DWs produce a positive, i.e., outward, particle flux in the positive $\eta_\alpha$ scenario. Only a weak negative, i.e., inward, particle flux appears, carried by high-$(k_y,k_\parallel)$ modes with weaker linear growth rates. Conversely, in the negative $\eta_\alpha$ scenario, a substantial negative particle flux is observed, predominantly carried by DWs.

To understand why the $E\times B$ particle flux is negative when $\eta_\alpha<0$, we examine the phase relation set by the linearized Ohm's law.
Starting from the linearized Ohm's law (Equation~(\ref{eq:linear_ohm})), the relation between density fluctuation $\tilde{n}$ and electrostatic potential fluctuation $\tilde{\phi}$ can be written as
\begin{equation}
    \label{eq:linear_ohm2}
    \frac{\tilde{n}}{n_0}=\frac{\tilde{\phi}}{\alpha_{de}T_{e0}}-\frac{\tilde{T}_e}{T_{e0}}+\frac{\alpha_m\eta_\parallel j_\parallel}{ik_\parallel\alpha_{de}T_{e0}}=\frac{\tilde{\phi}}{\alpha_{de}T_{e0}}\left(a+i\delta_t\right)
\end{equation}
where $\delta_t$ represents the relative phase shift between density fluctuation $\tilde{n}$ and electrostatic potential fluctuation $\tilde{\phi}$.
Consequently, the direction of $E\times B$ particle flux now solely depends on the sign of parameter $\delta_t$ as $\Gamma_n\propto-\delta_t|\tilde{\phi}|^2$.
For instance, a positive $\delta_t$, i.e., when $\tilde{n}$ leads $\tilde{\phi}$, results in a negative, or inward, particle flux, and vice versa.

Eliminating $\tilde{T}_e$ and $j_\parallel$ from Equation~(\ref{eq:linear_ohm2}) with the linearized electron temperature and density equations yields the temporal phase shift
\begin{equation}
    \label{eq:delta}
    \delta_t=\beta\left[-\alpha_T\left(\gamma+\chi\right)\left(1+\gamma\alpha_{R1}\right)+\alpha_{R2}\left(1+\gamma\alpha_{R1}\right)+\alpha_{R1}\omega_r\left(1+\alpha_T\omega_r\right)\right].
\end{equation}
Here, we separate the complex mode frequency $\omega=\omega_r+i\gamma$ and define the following parameters for simplicity:
\begin{equation}
    \begin{aligned}
        \alpha_T&=\frac{\alpha_{de}T_{e0}'k_y}{\omega_r^2+\left(\gamma+\chi\right)^2} \text{~with~}\chi=\kappa_\parallel^ek_\parallel^2,\quad \alpha_{R1}=\frac{\alpha_m\eta_\parallel n_0}{\epsilon_j\alpha_{de}T_{e0}k_\parallel^2}, \\
        \alpha_{R2}&=\frac{\alpha_m\eta_\parallel n_0'k_y}{\epsilon_jk_\parallel^2}, \quad \beta=\left[\left(1+\gamma \alpha_{R1}\right)^2+\omega_r^2\alpha_{R1}^2\right]^{-1}.
    \end{aligned}
\end{equation}

\begin{figure}[h!]
\centering
\includegraphics[width=0.5\textwidth]{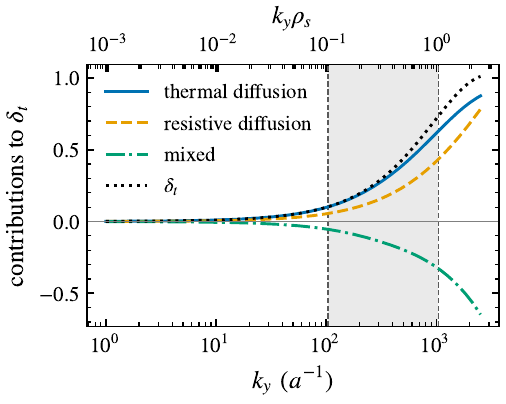}
\caption{The three terms of Equation~(\ref{eq:delta}) at the most unstable root along $k_\parallel R_0=0.7$, for $\eta_\alpha=-2$.}
\label{fig:delta_eta-2}
\end{figure}

Note that $\beta$, as well as $\chi,\alpha_{R1}$ are always positive, while the signs of $\alpha_{T}$ and $\alpha_{R2}$ depend on the local radial gradients of electron and density, respectively. Therefore, the sign of $\delta_t$ is determined by the three terms within the bracket on the right-hand-side of Equation~(\ref{eq:delta}). The first term in the bracket is always positive, producing an inward particle flux as long as $T_{e0}'<0$, i.e., the electron temperature decreases when moving radially outward. This inward particle flux, which is accompanied by the electron parallel thermal conduction process, i.e., $\alpha_T$ term, is often referred to as the electron thermal diffusion effects~\cite{coppi1978ion}.
The second term represents the resistive diffusion process and its sign depends on the background density gradient: it is positive for $n_0'>0$ and negative for $n_0'<0$.
The third term reflects mixed effects; its sign often depends on the direction of electron temperature gradient $T_{e0}'$ and hence is typically opposite to the sign of the first term. 
As illustrated in Figure~\ref{fig:delta_eta-2}, in the negative $\eta_\alpha$ scenario (e.g., for Figure~\ref{fig:lgr_eta-2}), the first two terms are both positive while the third is negative. Their sum is a positive phase shift $\delta_t$ that produces the inward particle flux in the area of interest on the spectrum (e.g., the high-$(k_y,k_\parallel)$ region of Figure~\ref{fig:lgr_eta-2}(b)). Averaging over the unstable modes within $k_y\rho_s\in(0.1,1)$ and weighting by their growth rates, the three terms contribute $+0.58$, $+0.76$ and $-0.23$ to $\delta_t=+1.12$.
The first two terms are comparable in this hollow-profile case but they have different fates. The resistive-diffusion term is proportional to $n_0'$ and vanishes as the density profile flattens, which is ordinary down-gradient diffusion and carries no pinch of its own once the gradient is gone. The thermal-diffusion term does not: repeating the same average for a flat density profile leaves $\delta_t=+0.35$, of which $+0.42$ comes from the $\alpha_T$ term alone. Electron thermal diffusion is therefore the process that keeps the turbulent flux inward once the hollow density gradient has been erased, which is the regime the simulation reaches at $t_2$.
In contrast, for the positive $\eta_\alpha$ scenario, the second term is negative and combines with the third term to overwhelm the first ($+0.32$, $-0.46$ and $-0.03$ under the same average, giving $\delta_t=-0.17$). Therefore, although thermal diffusion effects exist, they are unlikely to produce inward particle flux in this case.

\subsection{Verification of thermal-diffusion}

\begin{figure}[h!]
\centering
\includegraphics[width=\textwidth]{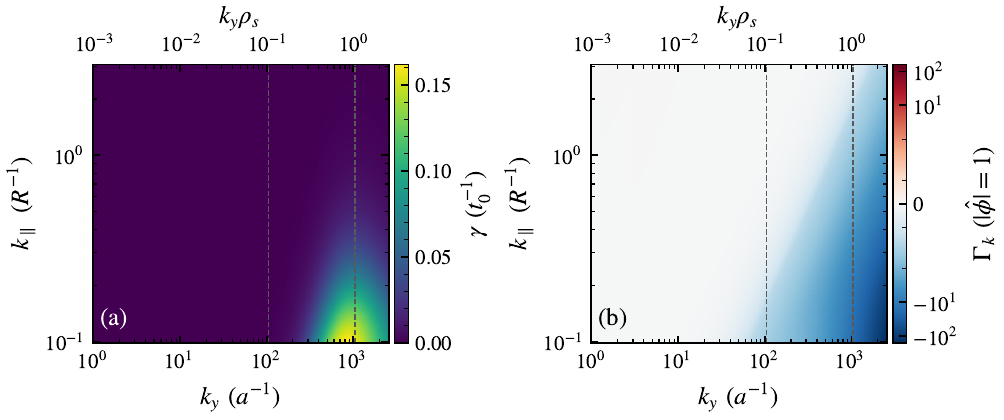}
\caption{Same setup as Figure~\ref{fig:lgr_eta-2}(a, b), but with the electron thermal response neglected.}
\label{fig:lgr_woTe}
\end{figure}

A direct check is to repeat the linear analysis without electron thermal fluctuations, i.e., $T_e=\text{constant}$. As illustrated in Figure~\ref{fig:lgr_woTe}, neglecting the electron thermal dynamics largely stabilizes the system: the maximum linear growth rate drops by a factor of $25$, from $3.9$ to $0.16\,t_0^{-1}$.  The inward $E\times B$ particle flux is thus likely negligible, although spectra evaluated at fixed mode amplitude cannot show that suppression directly; the nonlinear control simulation below confirms it.

\begin{figure}[h!]
\centering
\includegraphics[width=\textwidth]{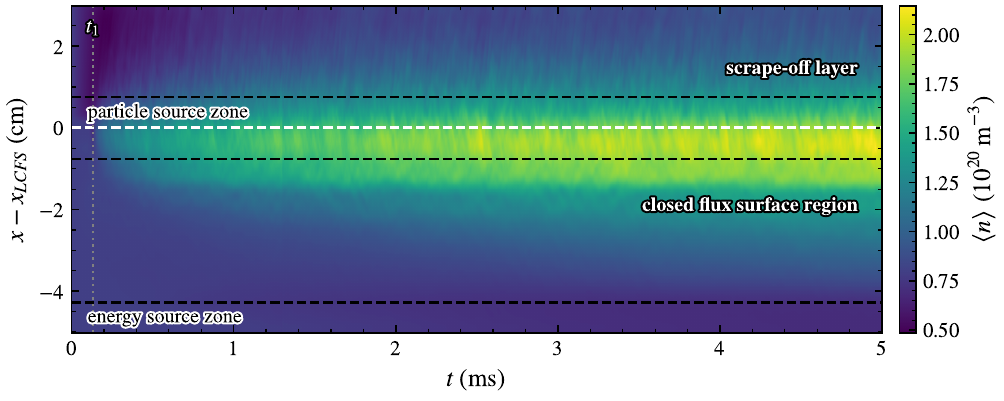}
\caption{Evolution of $\langle n \rangle$ similar as Figure~\ref{fig:dvt}, but with the $T_e$ equation turned off.}
\label{fig:den_evo_nt1c}
\end{figure}

\begin{figure}[h!]
\centering
\includegraphics[width=0.8\textwidth]{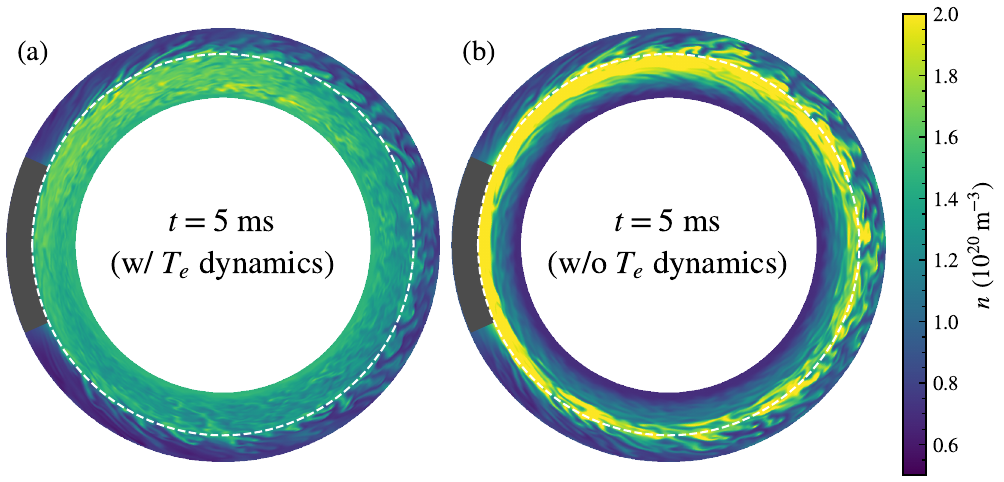}
\caption{Poloidal density cross-sections at $t=5$~ms (a) with and (b) without electron thermal dynamics.}
\label{fig:den_snapshots_comp}
\end{figure}

To verify these conclusions nonlinearly, we repeated the GDB simulation with the same setup but with electron thermal dynamics turned off ($\partial T_e/\partial t=0$). Figure~\ref{fig:den_evo_nt1c} shows the density-profile evolution of this control run, which lasted more than 5 ms, and Figure~\ref{fig:den_snapshots_comp} compares poloidal density snapshots of the two simulations at $t=5$ ms.

\begin{figure}[h!]
\centering
\includegraphics[width=0.5\textwidth]{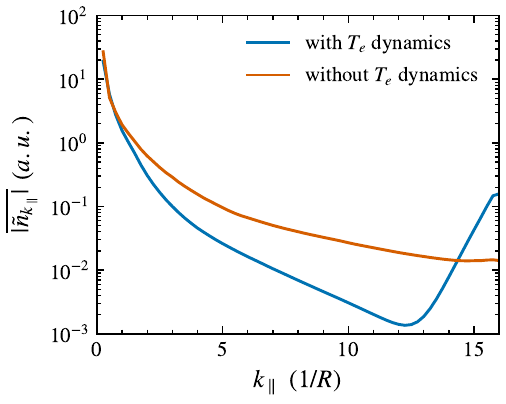}
\caption{Spatiotemporally averaged density fluctuation spectra at the $q=4$ surface (just outside the fueling region), with and without electron thermal dynamics.}
\label{fig:denk_par_comp}
\end{figure}

Without electron thermal dynamics, externally sourced plasma particles within the closed flux surface region are largely confined and accumulate near the particle fueling zone. This particle accumulation results in a steeper gradient and, consequently, stronger turbulence fluctuations; however, the inward particle flux remains small compared to the results of the original simulation with electron thermal dynamics. This weak inward flux, which advances the density front only slowly, is likely due to the inherent collisional and diffusive transport processes in the simulation (e.g., magnetic fluctuation and the hyper-diffusion included for numerical stability). 
Moreover, although it is difficult to distinguish between the two turbulence dynamics from the poloidal density snapshots (e.g., Figure~\ref{fig:den_snapshots_comp}) by eye, statistical analysis reveals that the turbulent fluctuation spectra differ markedly between the two simulations. As shown in Figure~\ref{fig:denk_par_comp}, without electron thermal dynamics, the density fluctuations are dominated by the lowest $k_\parallel$ mode, i.e., a characteristic of interchange-type instabilities such as RBMs, and the overall fluctuation level decreases monotonically as $k_\parallel$ increases. In contrast, in the simulation with electron thermal dynamics, a sub-dominant peak appears at higher $k_\parallel$ values, i.e., a feature of drift-wave turbulence, indicating the coexistence of two instabilities, RBMs and drift waves (DWs), in the original simulation.
  

Therefore, it is evident that electron thermal diffusion is largely responsible for driving the turbulent inward particle pinch when $\eta_\alpha<0$ in the simulation. Although the inward particle flux in this GDB simulation is carried mainly by drift-wave turbulence, the thermal diffusion process itself is more general and less sensitive to the type of turbulence present. 
Similarly, although the classical Spitzer-H\"arm expression for electron parallel thermal conductivity used in our simulations and many other codes~\cite{zholobenko2020thermal} is only valid for a collisional plasma, more sophisticated parallel heat flux models that extend into the weakly collisional parameter regime (e.g., flux-limited~\cite{zholobenko2024tokamak}, Landau fluid closures~\cite{zhu2021drift}, or kinetic closures~\cite{wang2019landau}) may impact the expression for $\chi$ but are unlikely to alter the overall picture.

\section{Edge density peaking at the late phase}
\label{sec:positive_eta}

With a good understanding of the inward particle flux in the negative $\eta_\alpha$ scenario, an even more intriguing question arises: what causes the edge density to further peak when $\eta_\alpha\geq 0$, i.e., after $t_2=4.15$ ms? This section investigates the mechanism of the edge density peaking observed during $t\in(4.15,8)$~ms. We organize the analysis around an explicit decomposition of the radial particle flux. Section~\ref{subsec:decomposition} introduces the averaging conventions, separates the contributions, and disposes of the diamagnetic term; Sections~\ref{subsec:classneo}--\ref{subsec:eqExBpinch} examine the remaining channels in turn. We find that the late-phase peaking is carried by the equilibrium $E\times B$ flux of the poloidally asymmetric density and potential. Classical neoclassical theory predicts these asymmetries, but the net $E\times B$ flux they drive cancels at leading order~\cite{hazeltine1973collision,helander2002collisional}. The flux we observe is the subleading remainder, made finite by non-ideal (resistive, inertial, and electromagnetic) corrections to the equilibrium electron force balance; our results therefore supplement, rather than contradict, the classical theory.

\subsection{Decomposition of the radial particle flux}
\label{subsec:decomposition}

In the drift-reduced Braginskii model, the radial particle flux contains an $E\times B$ contribution and a diamagnetic (magnetization) contribution. Under the large-aspect-ratio ordering adopted in this study the magnitude of the toroidal field is treated as constant in the perpendicular drift operators, so that no $\nabla B$ drift appears explicitly; the only magnetic-drift operator retained is the curvature operator $C(\cdot)$, whose associated perpendicular flow is the diamagnetic flow. Hence, in our simulation
\begin{equation}
    \label{eq:Gamma_decomp_drifts}
    \Gamma_{n,r} \;\simeq\; n V_{E,r} \;+\; n V_{\mathrm{dia},r},
\end{equation}
where $V_{E,r}\propto -\partial \phi/\partial y$ is the radial component of the $E\times B$ drift and $V_{\mathrm{dia},r}$ is the radial component of the diamagnetic flow.

To separate equilibrium from fluctuating contributions, we introduce two averages.  The \emph{time} average $\bar{f}\equiv (1/\Delta t)\int_{\Delta t}f\,dt$ is taken over a window $\Delta t$ long enough that turbulent fluctuations average to zero, i.e., $f(x,y,z,t)=\bar f(x,y)+\tilde f(x,y,z,t)$ with $\bar{\tilde f}\equiv 0$ and the $z$-independence of $\bar f$ follows from toroidal axisymmetry.  The \emph{flux-surface} average $\langle f\rangle$ (which, in our shift-circular geometry, is essentially a $y$- and $z$-average) further decomposes the equilibrium part into a poloidally symmetric component $\bar f_0(x)\equiv \langle \bar f\rangle$ and a poloidally asymmetric component $\bar f_1(x,y) \equiv \bar f - \bar f_0$, with $\langle \bar f_1\rangle = 0$ by definition.

Substituting these decompositions into Equation~(\ref{eq:Gamma_decomp_drifts}) and applying the combined time- and flux-surface average $\langle \overline{\,\cdot\,}\rangle$ to the full flux, the cross terms between $\bar f$ and $\tilde f$ vanish on time-averaging and the cross terms between $\bar f_0$ and $\bar f_1$ vanish on flux-surface averaging.  The result is
\begin{equation}
    \label{eq:Gamma_decomp_full}
    \langle \bar \Gamma_{n,r}\rangle \;=\;
      \underbrace{\bar \Gamma_{n,r}^{(\text{class.\,neo.})}}_{\text{(a)}}
      \;+\; \underbrace{\langle \tilde n\, \tilde V_{E,r}\rangle}_{\text{(b)}}
      \;+\; \underbrace{\langle \bar n_1\, \bar V_{E,r,1}\rangle}_{\text{(c)}}
      \;+\; \underbrace{\langle \bar n_1\, \bar V_{\mathrm{dia},r,1}\rangle}_{\text{(d)}},
\end{equation}
where term (a) groups all classical neoclassical contributions to the radial particle flux (e.g., parallel friction, ambipolar response, etc.), term (b) is the familiar turbulent cross-field $E\times B$ flux, term (c) is the equilibrium $E\times B$ flux carried by the poloidally asymmetric components, and term (d) is the corresponding equilibrium diamagnetic (magnetization) flux. Term (a) is the only channel classical neoclassical theory describes. Term (b) is fluctuation-driven transport beyond any equilibrium treatment. Term (c) cancels at leading order in the classical description, where the equilibrium electrons follow a Boltzmann response; Section~\ref{subsec:eqExBpinch} shows that it is finite here. Term (d) is not an independent transport channel at all, as shown next. 

We first dispose of term (d). The diamagnetic particle flux $n\gv{V}_{\mathrm{dia},\alpha}=(q_\alpha B)^{-1}\guv{b}_0\times\nabla p_\alpha$ is linear in the pressure (the density cancels), so its fluctuating part is $(q_\alpha B)^{-1}\guv{b}_0\times\nabla \tilde p_\alpha$ exactly and vanishes under the time average; this is why no turbulent diamagnetic term appears alongside term (b). Its equilibrium part, term (d), is a magnetization (curl) flow to leading order in the drift ordering. A magnetization flow transports no particles by itself: only its divergence enters the continuity equation, and that divergence reduces to the curvature compression $-2C(p_e)/(enB)$ already retained in the density equation of Section~\ref{sec:model}, and therefore already counted in the particle balance of Section~\ref{subsec:eqExBpinch}. 

Equation~(\ref{eq:Gamma_decomp_full}) also omits the electromagnetic ``flutter'' flux $n v_{\parallel i}\tilde{b}_r$, which is formally $O(\beta)\ll 1$ in the closed-flux region owing to the weak parallel flow~\cite{zhu2023electromagnetic}. In the late phase, however, it is inward and not negligible; being electromagnetic, it is another non-ideal channel, and its share of the density build-up is quantified in Section~\ref{subsec:eqExBpinch}.

\subsection{Classical neoclassical contribution}
\label{subsec:classneo}

Term (a) of Equation~(\ref{eq:Gamma_decomp_full}) is the focus of classical neoclassical theory.  In the Pfirsch--Schl\"uter regime relevant to our simulation ($\nu^*_e\simeq 10$--$300$ across the closed-flux analysis region; see the Summary), electron and ion temperature gradients alone do not drive a net radial particle pinch in the absence of an inductive toroidal electric field~\cite{hazeltine1973collision,helander2002collisional}.  The Ware pinch~\cite{ware1970pinch}, proportional to $E_\zeta$, is absent in our simulation by construction; the only other contributions to (a) are resistivity- and friction-induced terms which are generally small~\cite{helander2002collisional}.  We therefore expect $\bar\Gamma_{n,r}^{(\text{class.\,neo.})}\approx 0$ in our regime, consistent with classical theory; the late-phase pinch we report below cannot be attributed to this channel.

\subsection{Turbulent $E\times B$ channel}
\label{subsec:turbdiffusion}

Next we examine term (b), the turbulent $E\times B$ flux, which carried the inward pinch in the early phase via the electron thermal-diffusion process (Section~\ref{sec:negative_eta}). Once the density profile flattens, $L_n\to\infty$ and the thermal-diffusion drive weakens. The linear analysis of Section~\ref{sec:negative_eta} already quantifies that weakening: with $\Gamma_n\propto-\delta_t|\tilde\phi|^2$, the flux carried per unit fluctuation amplitude follows $\delta_t$, which falls from $+1.12$ at $\eta_\alpha=-2$ to $+0.35$ for a flat profile (Figure~\ref{fig:lgr_eta_inf}). 
A particle-balance analysis of the nonlinear simulation confirms this transition: evaluated window by window, it shows that the turbulent $E\times B$ channel supplies about half of the observed density build-up during the early phase, but that its net contribution across the closed-flux analysis region reverses sign (net outward) beyond $t_2$ once $\eta_\alpha>0$. 
The turbulent $E\times B$ channel alone is therefore insufficient to explain the late-phase density peaking.

\begin{figure}[h!]
\centering
\includegraphics[width=\textwidth]{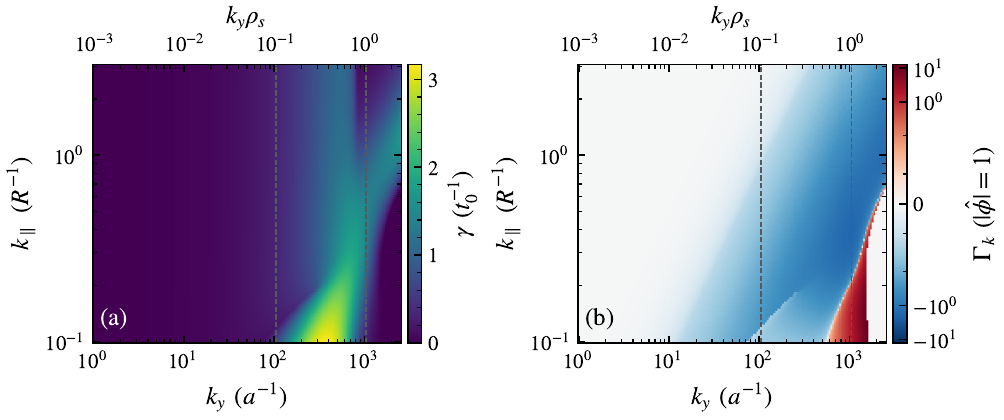}
\caption{Same setup as Figure~\ref{fig:lgr_eta-2}(a, b), but for a flat density profile ($L_n=\infty$).}
\label{fig:lgr_eta_inf}
\end{figure}

\subsection{The non-ideal equilibrium $E\times B$ pinch}
\label{subsec:eqExBpinch}

Having shown that contributions (a) and (b) cannot account for the late-phase peaking, we turn to contribution (c), the equilibrium $E\times B$ flux driven by the poloidally asymmetric components $\bar n_1$ and $\bar V_{E,r,1}$.  A notable feature of the simulation, already documented in previous work~\cite{zhu2018up}, is a substantial up-down asymmetric density profile, as shown in the poloidal density snapshots (e.g., Figures~\ref{fig:den_snapshots_peaking}(c-d), \ref{fig:den_snapshots_comp}(a)).  This up-down symmetry breaking is induced by the binormal component of the Pfirsch--Schl\"uter ion heat flux $\gv{q}_{i\wedge}$, which makes the equilibrium ion temperature $\bar T_{i,1}$ up-down asymmetric; through the force-balance constraint, the equilibrium density profile inherits a corresponding asymmetric component $\bar n_1$. 

\begin{figure}[h!]
\centering
\includegraphics[width=0.8\textwidth]{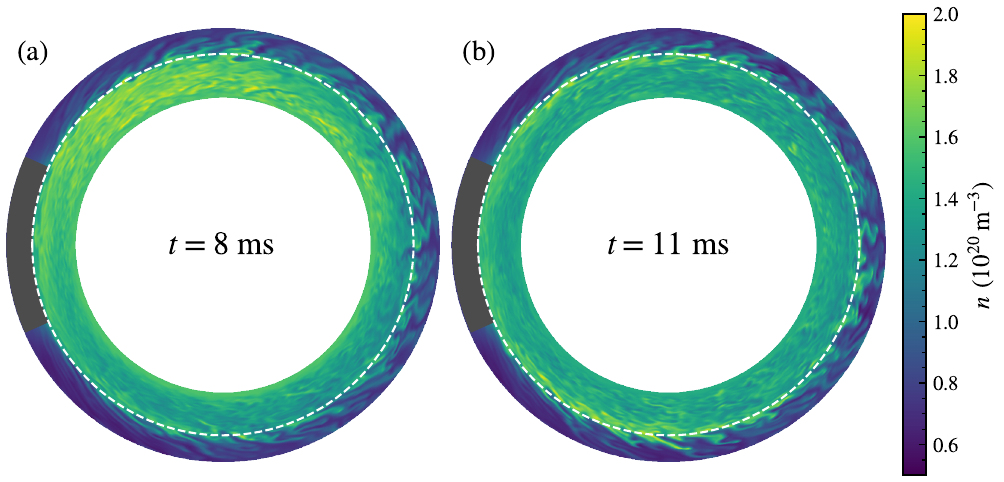}
\caption{Poloidal density cross-sections (a) at peak, $t=8$~ms, and (b) after relaxation, $t=11$~ms.}
\label{fig:den_snapshots_relaxation}
\end{figure}

To verify that the up-down asymmetric profile is the proximate cause of the late-phase peaking, we continued the simulation for another $4$~ms with the transverse heat flux terms (the $C(T_\alpha)$ terms) turned off in the temperature equations.  As Figure~\ref{fig:dvt2} shows, the centrally peaked density profile begins to relax as soon as these terms are off, and by $t_4\simeq 11$~ms a nearly flat profile is re-established in the closed-flux region outside the fueling zone.  The poloidal snapshots at $t=8$ and $11$~ms in Figure~\ref{fig:den_snapshots_relaxation} show the density becoming up-down symmetric, confirming that the late-phase peaking is tied to the up-down asymmetric $\bar n_1$.

\begin{figure}[h!]
\centering
\includegraphics[width=\textwidth]{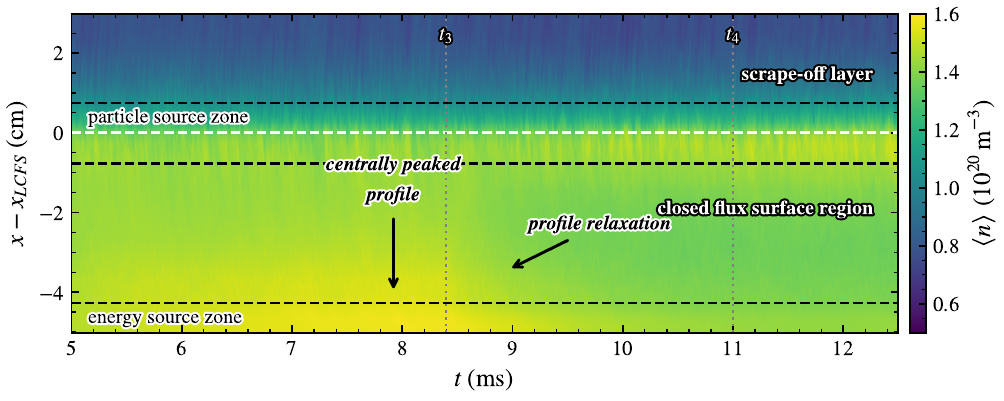}
\caption{Continuation of the simulation of Figure~\ref{fig:dvt} with the transverse heat flux terms turned off at $t_3=8.40$~ms; $\langle n\rangle$ reaches a new quasi-steady state by $t_4\simeq11$~ms.}
\label{fig:dvt2}
\end{figure}

This poloidal symmetry breaking chain is genuinely neoclassical and is captured within the classical Pfirsch--Schl\"uter theory. Classical neoclassical theory finds no net equilibrium $E\times B$ radial flux from these asymmetries at leading order, as discussed in Section~\ref{subsec:classneo}, so the inward flux observed in our simulation and analyzed below must be the subleading remainder.


Whether term (c) drives a non-zero net radial flux is a question about the equilibrium phase relationship between $\bar n_1$ and $\bar \phi_1$.  In the classical neoclassical limit, the equilibrium electron parallel force balance reduces to a Boltzmann response (often called the adiabatic electron response),
\begin{equation}
    \label{eq:adiabatic_limit}
    \frac{\bar n_1}{n_0} \;=\; \frac{e\bar\phi_1}{T_{e0}},
\end{equation}
i.e., $\bar n_1$ and $\bar\phi_1$ are exactly in phase poloidally. In that case $\bar V_{E,r,1}\propto -\partial \bar\phi_1/\partial y$ is in quadrature with $\bar n_1$, so $\langle \bar n_1\,\bar V_{E,r,1}\rangle = 0$.  In the collisional analysis~\cite{hazeltine1973collision} this relation is not an assumption but the leading-order result. The electron kinetic analysis there yields precisely the Boltzmann variation of the electron density within a flux surface, and the correction from parallel friction alone is $O(\sqrt{m_e/m_i})$ smaller. The equilibrium $E\times B$ channel thus vanishes at leading order. 
The generalized Ohm's law in GDB contains more non-ideal physics ingredients: it retains parallel resistivity $\eta_\parallel j_\parallel$, electron parallel inertia (the $d_e^2$ term), and electromagnetic induction $\partial \psi^*/\partial t$, all of which contribute to the equilibrium electron force balance for the asymmetric components and break Equation~(\ref{eq:adiabatic_limit}). The result is a small but finite poloidal phase shift $\delta_s$ between $\bar n_1$ and $\bar\phi_1$, and hence a finite equilibrium $E\times B$ flux at higher order.

\begin{figure}[h!]
\centering
\includegraphics[width=0.8\textwidth]{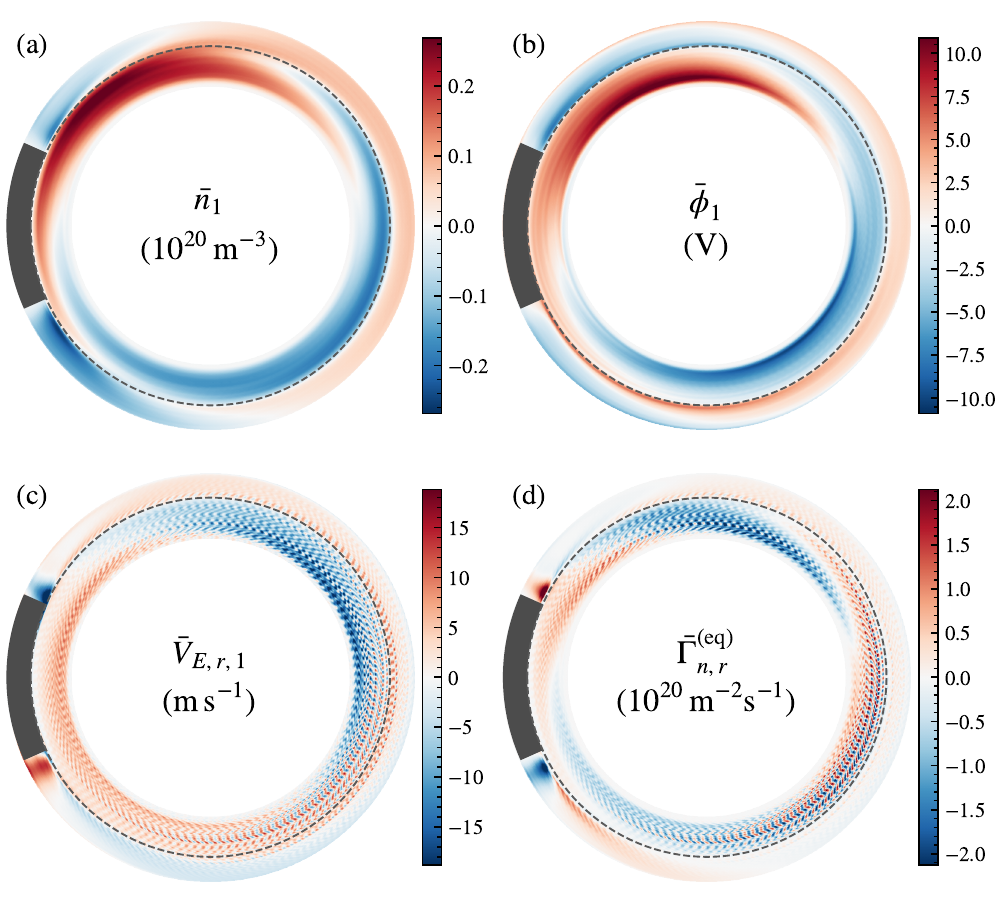}
\caption{Poloidally asymmetric components at $t\in(8,8.4)$~ms: (a) $\bar{n}_1$, (b) $\bar{\phi}_1$, (c) $\bar{V}_{E,r,1}$, and (d) the equilibrium flux $\bar{\Gamma}_{n,r}^{(\text{eq})}=\bar n_1\,\bar V_{E,r,1}$.}
\label{fig:gamman}
\end{figure}

Figure~\ref{fig:gamman} illustrates the poloidal asymmetric components $\bar n_1$, $\bar\phi_1$, $\bar V_{E,r,1}$ and the resulting equilibrium flux $\bar \Gamma_{n,r}^{(\text{eq})}\equiv \bar n_1\,\bar V_{E,r,1}$ in the late phase, specifically for $t\in(8,8.4)$~ms. Both $\bar n_1$ and $\bar\phi_1$ show clear up-down asymmetric structure of order $10\%$ of their flux-surface-averaged values, and the resulting $\bar V_{E,r,1}$ and $\bar\Gamma_{n,r}^{(\text{eq})}$ exhibit consistent poloidal variation.

\begin{figure}[h!]
\centering
\includegraphics[width=\textwidth]{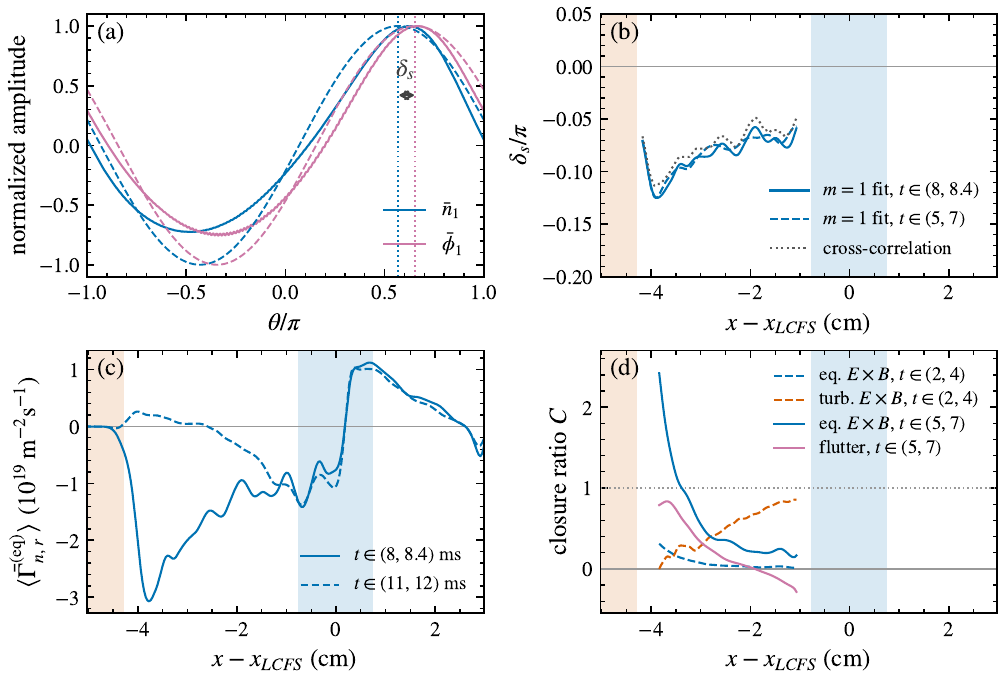}
\caption{(a) Normalized $\bar{n}_1$ and $\bar{\phi}_1$ at $x-x_\text{LCFS}=-3$~cm over $t\in(8,8.4)$~ms, with $m{=}1$ harmonic fits (dashed) and peak locations (dotted); the arrow marks the peak offset $\delta_s$.  (b) $\delta_s(x)$ from the $m{=}1$ fit for $t\in(8,8.4)$ and $(5,7)$~ms, and from the shift-of-best-overlap measurement, across the closed-flux analysis region.  (c) the equilibrium flux $\langle\bar\Gamma^{(\text{eq})}_{n,r}\rangle$ with (solid) and without (dashed, $t\in(11,12)$~ms) the transverse heat flux terms.  (d) the closure ratio $C(r)$ over the early ($t\in(2,4)$~ms, dashed) and peaking ($t\in(5,7)$~ms, solid) windows; the dotted line marks full closure, $C=1$.  The flutter curve is shown for the peaking window only, where the cleaned $\tilde b_r$ data exist, and the turbulent curve for the early window only (see text).}
\label{fig:delta_s}
\end{figure}

Following the same analysis performed in the previous study~\cite{zhu2018up}, the up-down asymmetric density driven by Pfirsch--Schl\"uter ion heat flux can be written as $\bar{n}_1=|\bar{n}_1|\sin\theta$, compensating $\bar{T}_{i,1}\propto -\sin\theta$. Here $\theta\in[-\pi,\pi)$ is the poloidal angle, measured from the outboard midplane and increasing counter-clockwise (so that the inner-wall limiter sits at $\theta=\pm\pi$). In the Boltzmann limit of Equation~(\ref{eq:adiabatic_limit}), $\bar{\phi}_1$ follows the same poloidal variation as $\bar{n}_1$, i.e., $\bar{\phi}_1=|\bar{\phi}_1|\sin\theta$, and the resulting equilibrium particle flux $\bar\Gamma_{n,r}^{(\text{eq})}\propto |\bar{n}_1||\bar{\phi}_1|\sin\theta\cos\theta$ vanishes after flux-surface averaging. With a non-Boltzmann correction $\bar{\phi}_1= |\bar\phi_1|\sin\left(\theta+\delta_s\right)$, however, writing the radial drift explicitly as $\bar V_{E,r,1}=-(rB)^{-1}\partial\bar\phi_1/\partial\theta$,
\begin{equation}
    \label{eq:GammaEB_sin}
    \langle \bar\Gamma_{n,r}^{(\text{eq})}\rangle \;=\; -\frac{|\bar n_1|\,|\bar\phi_1|}{rB}\,\langle \sin\theta\cos(\theta+\delta_s)\rangle \;=\; \frac{|\bar n_1|\,|\bar\phi_1|}{2rB}\,\sin\delta_s,
\end{equation}
which is inward ($<0$) for $\delta_s\in(-\pi,0)$, i.e., when $\bar{n}_1$ lags $\bar{\phi}_1$ poloidally. Figure~\ref{fig:delta_s}(a) shows that this phase shift is indeed present in the simulation: an $m{=}1$ harmonic fit of $\bar n_1$ and $\bar\phi_1$ across the closed-flux analysis region during $t\in(8,8.4)$~ms yields $\delta_s\simeq-0.08\pi$ (median), with a $10$--$90\%$ spread of $(-0.11,-0.07)\pi$ and nearly identical values during $t\in(5,7)$~ms (Figure~\ref{fig:delta_s}(b)). As a check that this offset is not an artifact of assuming a sinusoidal shape, a mode-independent, direct cross-poloidal-profile fit is performed and yields $\delta_s\simeq-0.07\pi$, within $0.01\pi$ of the harmonic fit. Through Equation~(\ref{eq:GammaEB_sin}), this phase shift produces the inward $\bar\Gamma_{n,r}^{(\text{eq})}$ plotted in Figure~\ref{fig:delta_s}(c). In the relaxation continuation ($t>t_3=8.4$~ms, dashed lines in Figure~\ref{fig:delta_s}(c)), the up-down asymmetric components $\bar{n}_1$ and $\bar{\phi}_1$ decay together and the equilibrium inward flux essentially vanishes across most of the closed-flux region; the residual localized flux near the source zone is associated with a different, in-out symmetry breaking discussed at the end of this subsection.

To test whether the equilibrium $E\times B$ contribution accounts \emph{quantitatively} for the central density build-up, we compare the particle flow it carries with the observed source-corrected accumulation.  For the shell between the outer edge of the energy-source zone, $r_E$, and a radius $r$ inside the particle-source zone, the equilibrium net inflow and the observed accumulation are
\begin{equation*}
    \Phi_{\text{eq}}(r) \;\equiv\; A(r)\,\langle \bar n_1\,\bar V_{E,r,1}\rangle, \qquad \dot N_{\text{obs}}(r) \;\equiv\; \int_{r_E<r'<r}\!\left[\frac{\partial \langle n\rangle}{\partial t} \;-\; \langle S_n\rangle \right] dV,
\end{equation*}
with the contribution, or closure, ratio
\begin{equation*}
    C(r) \;=\; -\left[\Phi_{\text{eq}}(r)-\Phi_{\text{eq}}(r_E)\right]\big/\dot N_{\text{obs}}(r),
\end{equation*}
which measures the fraction of the observed source-corrected accumulation in the shell that is supplied by the net equilibrium $E\times B$ inflow through its boundaries: $C(r)=1$ means this channel alone accounts for the entire density build-up between $r_E$ and $r$, while $C\ll 1$ marks a negligible contribution.  Here $A(r)=4\pi^2 r R_0$ is the flux-surface area.  Because the density build-up occurs during the peaking phase (the profile is quasi-stationary by $t\approx 8$~ms), the contribuation is evaluated over $t\in(5,7)$~ms.  Figure~\ref{fig:delta_s}(d) shows $C(r)$ for each channel.  We quote $C$ quantitatively for the early window, where the $E\times B$ particle balance closes: over $t\in(2,4)$~ms the divergence of the $E\times B$ flux accounts for the observed accumulation to within a few percent, and $C$ then shows the turbulent channel supplying about half of it ($0.50$) against a negligible $0.04$ for the equilibrium channel.  This is the quantitative counterpart of Section~\ref{sec:negative_eta}: in the early phase the pinch is turbulent.
By the peaking window the picture has reversed.  The equilibrium share has grown to $0.36$ (region median), rising with depth from $\approx0.2$ near the fueling side to $\approx1.2$ at the innermost analysis radii, while the net turbulent contribution across the region has changed sign to outward.  These late-window numbers are indicative rather than a closed budget, for two reasons.  The density profile is approaching its steady state, so the net accumulation has fallen by nearly an order of magnitude and $C$ is a ratio to a small residual; and a $2$~ms time average no longer separates the fluctuating flux cleanly from the equilibrium still drifting through the window, so the $E\times B$ balance over $t\in(5,7)$~ms does not close as it does over $t\in(2,4)$~ms.  Equation~(\ref{eq:Gamma_decomp_full}) also omits the magnetization compression and the hyper-diffusive channel, so the curves are not expected to sum to unity in any window.
The case for the late phase therefore rests not on $C$ alone but on the sign of the equilibrium flux, on its growth with depth and with time, and on the relaxation test of Figure~\ref{fig:delta_s}(c).  Together these identify the equilibrium $E\times B$ channel as the dominant $E\times B$ carrier of the late-phase pinch.
The magnetic-flutter contribution set aside in Section~\ref{subsec:decomposition} adds to this rather than competing with it: over the same window it is also inward, with roughly half the magnitude of the equilibrium $E\times B$ term, supplying a further $\simeq0.2$ of the observed build-up.  Carried by the fluctuating radial field $\tilde b_r$, it is also non-ideal, another channel absent from the leading-order classical description, so the late-phase inward flux rests on two non-ideal channels rather than one.

The above analysis establishes the late-phase pinch as a \emph{non-ideal equilibrium $E\times B$ flux}.  It exists because the non-ideal Ohm's-law terms sustain a small but finite departure from the Boltzmann relation, in the form of the poloidal phase shift $\delta_s$. Classical neoclassical theory predicts no net radial flux from this poloidal symmetry breaking, because at leading order the electrons follow the Boltzmann relation; our result shows that a subleading flux appears once Equation~(\ref{eq:adiabatic_limit}) is broken. It therefore supplements, rather than contradicts, the classical theory.

The phase shift $\delta_s$ used here must not be confused with $\delta_t$ of Section~\ref{sec:negative_eta}.  $\delta_s$ represents a spatial (poloidal) phase shift between two time-stationary quantities $\bar{n}_1$ and $\bar{\phi}_1$, characterizing the equilibrium asymmetry; $\delta_t$ represents a temporal phase shift between two fluctuating quantities $\tilde{n}$ and $\tilde{\phi}$, characterizing the turbulent transport.

Finally, despite the radial density profile relaxing in the simulation once the transverse heat flux terms are turned off after $t=8.4$~ms, further analysis found that a localized inward particle flux still persists near the particle source zone, i.e., roughly $x-x_{LCFS}\in(-2,-1)$~cm as shown in Figure~\ref{fig:delta_s}(c).  This phenomenon occurs because ballooning-type turbulence, which favors the outboard side, can produce an in-out asymmetric plasma.  Consequently, this poloidal symmetry breaking leads to finite values of $\bar{n}_1$ and $\bar{\phi}_1$, along with a finite $\delta_s$ through the same non-ideal and nonlinear processes.  A negative equilibrium particle flux is therefore possible.  Nevertheless, in this study, the in-out-asymmetric inward particle flux is localized and too weak to produce a centrally peaked density profile.

\section{Summary}
\label{sec:summary}

We performed a flux-driven global edge turbulence simulation to study inward particle transport in the tokamak edge region. Unlike most edge turbulence studies, which fix the core-side density or fuel the plasma from the core and follow sub-millisecond dynamics, this simulation starts from a flat radial density profile, fuels the plasma locally near the last closed flux surface, and runs over a transport time scale ($O(10)$ ms) to a quasi-steady-state density profile. 
A robust inward particle flux appears in the closed-flux region away from the fueling zone and eventually builds a centrally peaked density profile. Two distinct pinch mechanisms carry it. In the early stage, while the density gradient is positive (a hollow or edge-humped profile), linear analysis and a controlled simulation show that the fluctuating $E\times B$ convection of drift-wave turbulence carries the inward particle flux, alongside outward heat flux, through the electron thermal-diffusion process. This mechanism weakens as the density profile flattens ($1/L_n\to 0$) and cannot alone explain the late-stage peaking. 
Further analysis identifies the second mechanism as a \emph{non-ideal equilibrium $E\times B$ flux}.  The Pfirsch--Schl\"uter ion heat flux drives an up-down asymmetric ion temperature and, through pressure balance, an up-down asymmetric equilibrium density $\bar n_1$; this drive is genuinely neoclassical. Classical theory, however, finds no net radial flux from these asymmetries: in the strict classical limit the equilibrium electrons follow the Boltzmann relation, so $\bar n_1$ and $\bar\phi_1$ are exactly in phase, and the equilibrium $E\times B$ flux $\langle \bar n_1 \bar V_{E,r,1}\rangle$ cancels identically, which is the classical no-pinch result in the Pfirsch--Schl\"uter regime~\cite{hazeltine1973collision}.  In GDB, parallel resistivity, electron parallel inertia, and electromagnetic induction in Ohm's law break the Boltzmann relation and produce a small poloidal phase shift $\delta_s\simeq-0.08\pi$ on average across the closed-flux analysis region. This phase shift yields a finite inward flux that supplies the observed late-phase density build-up and supplements rather than contradicts the classical theory.  Although the two pinch mechanisms, one carried by turbulent fluctuations and the other by equilibrium asymmetries with a non-Boltzmann electron response, are fundamentally different, they coexist throughout the simulation, and their relative importance varies with the local density gradient.

The collisional (Braginskii) closure underlying GDB is quantitatively justified for this case: evaluated with the steady-state profiles, the normalized collisionality across the closed-flux analysis region is $\nu^*_e\simeq 9$--$310$ and $\nu^*_i\simeq 6$--$110$, at or above the Pfirsch--Schl\"uter threshold $\epsilon^{-3/2}\simeq 6$ everywhere, with only the hottest core-side edge approaching the plateau boundary.  Such parameters are characteristic of high-density, low-temperature (C-Mod-like) edges; hotter and less collisional edge plasmas, including large portions of projected reactor edges, fall outside this regime, and there a transcollisional closure~\cite{zholobenko2024tokamak} becomes necessary.

Although this study uses a shift-circular tokamak edge configuration, the pinch mechanisms generalize: the equilibrium $E\times B$ pinch requires only poloidal symmetry breaking and a non-Boltzmann equilibrium electron response, ingredients also present in stellarator edges and in tokamaks with applied resonant magnetic perturbations. These insights into edge particle transport bear directly on density-profile prediction, pedestal formation, and core-edge integration.

\begin{acknowledgments}
The author would like to thank Drs. Andreas Stegmeir, Manaure Francisquez, and Sarah Rogers for helpful discussions.
This work was performed under the auspices of the U.S. Department of Energy (DOE) by Lawrence Livermore National Laboratory (LLNL) under Contract DE-AC52-07NA27344. This research used resources of the National Energy Research Scientific Computing Center, a DOE Office of Science User Facility supported by the Office of Science of the U.S. Department of Energy under Contract No. DE-AC02-05CH11231 (FES-ERCAP 0036750). LLNL-JRNL-2003017
\end{acknowledgments}

\section*{Data Availability Statement}
The data that support the findings of this study are available from the corresponding author upon reasonable request.

\appendix
\section{Normalization of drift-Braginskii equation in GDB}\label{app:normalization}

In the GDB model, the drift-reduced Braginskii equations were normalized to a dimensionless form prior to simulations. The perpendicular and parallel characteristic lengths, $L_\perp$ and $L_\parallel$, are chosen as the on-axis minor and major radii, $a$ and $R_0$, respectively. The reference time is defined as $t_0=\sqrt{aR_0/2}/c_{s0}$ where the reference sound speed is given by $c_{s0}=\sqrt{T_0/m_i}$ based on the reference temperature $T_0$. Additionally, the magnetic field strength, plasma density, electrostatic potential, and parallel magnetic potential are normalized to $B_0$, $n_0$, $B_0a^2/ct_0$ and $a\beta_0B_0$, respectively, where $\beta_0=8\pi n_0T_0/B_0^2$.

Under these normalizations, the actual set of equations that evolves is:
\begin{align}
    \frac{\partial \ln n}{\partial t}&=-\left[\phi,\ln n\right]-\epsilon_v v_{\parallel i}\nabla_\parallel \ln n+\epsilon_R\left(C(\phi)-\alpha_d\frac{C(p_e)}{n}+\alpha_d\frac{\nabla_\parallel j_\parallel}{n}\right) -\epsilon_v\nabla_\parallel v_{\parallel i}+\frac{S_n}{n},\\
    \frac{\partial \ln T_e}{\partial t}&=-\left[\phi, \ln T_e\right]-v_{\parallel e}\nabla_\parallel \ln T_e+\frac{\kappa_\parallel^\text{e}}{nT_e}\nabla_\parallel \left(T_e^{7/2}\nabla_\parallel \ln T_e\right)-\frac{5}{3}\epsilon_R\alpha_d T_e C(\ln T_e) \nonumber \\
    &+\frac{2}{3}\left[\epsilon_R\left(C(\phi)-\alpha_d\frac{C(p_e)}{n}+\alpha_d\frac{\nabla_\parallel j_\parallel}{n}\right) -\epsilon_v\nabla_\parallel v_{\parallel i}-\epsilon_R \alpha_d\frac{j_\parallel}{n}\nabla_\parallel \ln n\right]+\frac{S_{T_e}}{T_e},\\
    \frac{\partial \ln T_i}{\partial t}&=-\left[\phi,\ln T_i\right]-\epsilon_v v_{\parallel i}\nabla_\parallel \ln T_i+\frac{\kappa_\parallel^\text{i}}{nT_i}\nabla_\parallel T_i^{7/2}\left(\nabla_\parallel \ln T_i\right)+\frac{5}{3}\epsilon_R\alpha_d T_iC(\ln T_i) \nonumber \\
    &+\frac{2}{3}\left[\epsilon_R \left(C(\phi)-\alpha_d\frac{C(p_e)}{n}+\alpha_d\frac{\nabla_\parallel j_\parallel }{n}\right) -\epsilon_v\nabla_\parallel v_{\parallel i}\right]+\frac{S_{T_i}}{T_i},\\
    \frac{\partial \varpi}{\partial t}&=-C(p_e+p_i)+\nabla_\parallel j_\parallel-\epsilon_G C(G_i) -\nabla_\perp\cdot\left(n\left[\phi,\nabla_\perp\phi+\alpha_d\frac{\nabla_\perp p_i}{n}\right]\right),\\
    \frac{\partial v_{\parallel i}}{\partial t}&=-\left[\phi,v_{\parallel i}\right]-\epsilon_v v_{\parallel i}\nabla_\parallel v_{\parallel i}-\epsilon_v\left[\frac{\nabla_\parallel \left(p_e+p_i\right)}{n}-4\epsilon_G\frac{\nabla_\parallel G_i}{n}\right]+\epsilon_R \alpha_d T_iC(v_{\parallel i}),\\
    \frac{\partial \psi^*}{\partial t}&=d_\text{e}^2\left[\phi,\frac{j_\parallel}{n}\right]+d_\text{e}^2v_{\parallel e}\nabla_\parallel\left(\frac{j_\parallel}{n}\right)+\frac{1}{\alpha_m}\left(\nabla_\parallel \phi-\alpha_d\frac{\nabla_\parallel p_e}{n}\right)+\frac{\eta_\parallel}{T_e^{3/2}} j_\parallel,
\end{align}
with
\begin{align}
    v_{\parallel e}&=\epsilon_vv_{\parallel i}-\epsilon_R\alpha_d \frac{j_\parallel}{n}, \\
    \varpi&=\nabla_\perp\cdot\langle n\rangle\left(\nabla_\perp\phi+\alpha_d\frac{\nabla p_i}{n}\right),\\
    G_i&=T_i^{5/2}\left[\epsilon_R\left(C(\phi)+\alpha_d\frac{C(p_i)}{n}\right)-4\epsilon_v\nabla_\parallel v_{\parallel i}\right].
\end{align}

The normalization coefficients are defined as follows:
\begin{equation}
    \begin{aligned}
        \epsilon_v&=c_{s0}t_0/R_0,\quad \epsilon_R=2a/R_0,\quad \alpha_d=\frac{c_{s0}^2t_0}{\omega_{ci}a^2},\quad \epsilon_G=0.08\frac{\tau_{i0}}{t_0},\quad \alpha_m=\frac{R_0\beta_0}{a},\\
        \kappa_\parallel^\alpha&=2h_\alpha\frac{t_0\tau_{\alpha0}T_0}{R_0^2m_\alpha}\text{(~with~}h_e\simeq3.9,~h_i\simeq3.2),\quad d_\text{e}^2=\left(\frac{c}{a\omega_{pe0}}\right)^2,\quad \eta_\parallel=0.51\frac{t_0d_\text{e}^2}{\tau_{e0}}.
    \end{aligned}
\end{equation}
Here, $\tau_{i0}, \tau_{e0}$ and $\omega_{pe0}$ represent the reference ion and electron collisional times, as well as the plasma frequency, respectively.


For numerical stability reasons, a small amount of sixth-order anisotropic hyper-diffusion, i.e., $\nu\nabla_\perp^6 f$, without mixed-derivative terms is applied to each equation to dissipate grid-scale turbulent structures, as noted in \cite{francisquez2019multigrid}. In this study, the value of $\nu$ is on the order of $10^{-15}$ in the normalized units along the radial direction and $10^{-12}$ along the poloidal direction.

\bibliography{dp}

\end{document}